\documentclass{aa}
\usepackage{graphicx,epsfig}
\begin{document}

\newcommand{\ea}{et al.}
\newcommand{\beq}{\begin{equation}}
\newcommand{\enq}{\end{equation}}
\newcommand{\bfg}{\begin{figure}}
\newcommand{\efg}{\end{figure}}
\newcommand{\bfa}{\begin{figure*}}
\newcommand{\efa}{\end{figure*}}
\newcommand{\bea}{\begin{eqnarray}}
\newcommand{\ena}{\end{eqnarray}}  
\def\apj{Astroph. J.}
\def\apjl{Astroph. J. Lett.}
\def\asa{A\&A}
\def\mnras{MNRAS}
\def\phr{Phys. Rev.}
\def\phrl{Phys. Rev. Lett.}
\def\ltsima{$\; \buildrel < \over \sim \;$}
\def\gtsima{$\; \buildrel > \over \sim \;$}
\def\simlt{\lower.5ex\hbox{\ltsima}}
\def\simgt{\lower.5ex\hbox{\gtsima}}
\def\ebf{{\bf e}}
\def\Ebf{{\bf E}}
\def\dpar{{\bf \partial}}    
\def\ka{{\rm K}$\alpha$}
\def\kb{{\rm K}$\beta$} 

\title{On the origin of the broad, relativistic iron line of MCG--6-30-15
observed by {\fontfamily{ptm}\fontshape{it}\selectfont XMM-Newton}}
\author{ Andrea Martocchia\inst{1}, Giorgio Matt\inst{1},  
and Vladim{\'\i}r Karas\inst{2} }
\institute{       
Dipartimento di Fisica, Universit\`a degli Studi "ROMA TRE", 
Via della Vasca Navale 84, I--00146 Roma (Italy)\\
\email{martocchia@fis.uniroma3.it, matt@fis.uniroma3.it}
\and   
Astronomical Institute of the Charles University, 
Faculty of Mathematics and Physics, 
V Hole\v{s}ovi\v{c}k\'ach 2, CZ--180 00 Praha (Czech Republic)\\
\email{vladimir.karas@mff.cuni.cz}     
}

\date{Received... Accepted...}


\abstract{The relativistic iron line profile recently observed
by {\it XMM-Newton} in the spectrum of the Seyfert 1 galaxy MCG--6-30-15
(Wilms et al., 2001) is discussed in the framework of
the {\it lamp-post\/} model.
It is shown that the steep disc emissivity, the large line
equivalent width and the amount of Compton reflection can be
self-consistently reproduced in this scenario.}

\authorrunning{A. Martocchia, G. Matt \& V. Karas}
\titlerunning{On the broad, relativistic iron line of MCG--6-30-15 }
\maketitle


\section{Introduction}

Wilms et al. (2001; cited as W01 hereafter) recently presented 
and discussed an extremely broad and red-shifted 
iron K$\alpha$ feature detected in the 06/11-12/2000 100 ksec 
{\it XMM-Newton} observation of MCG--6-30-15. This Seyfert 1 galaxy
is well known for possessing one of the best studied broad iron lines,
whose profile is explained by relativistic effects (Tanaka \ea, 1995, 
Guainazzi \ea, 1998; see Fabian \ea, 2000, for a review). 
The Fe \ka\ profile observed by {\it XMM-Newton}'s {\it EPIC-pn} camera 
is similar to the one
observed by Iwasawa et al. (1996) using {\it ASCA} data 
during a short ($\sim 15.2$ ksec) period of low X-ray flux. 
W01 may have caught the source in a similar ``deep minimum state", 
i.e. a state in which
the primary flux is lower ($F_{\rm 2-10 ~keV}=2.3 \times 10^{-11}$ 
erg s$^{-1}$ cm$^{-2}$) and the line Equivalent Width (EW) higher 
(up to $300\div400$ eV) than the time-averaged values. 

The line profile indicates that a large fraction of the emission 
comes from $r<6r_{\rm{g}}$ ($r_{\rm{g}}=m={GM \over c^2}$).
This implies either that the
central Black Hole (BH) is rotating -- thus the radius of the
disc innermost stable orbit, $r_{\rm{ms}}$, lies between $6r_{\rm{g}}$ 
(=$r_{\rm{ms}}$ 
for a static BH) and $1.23r_{\rm{g}}$ ($r_{\rm{ms}}$ 
of a canonically spinning BH: 
Thorne, 1974) -- or that the fluorescent line emission originates from 
matter falling freely below $r_{\rm{ms}}$ (Reynolds \& Begelman, 1997).
Sako et al. (2001) proposed that also some spectral features observed
at lower energies in this source, as well as in Mkn 766, are 
Ly$\alpha$ lines of carbon, nitrogen, and oxygen, affected by relativistic 
broadening in the spacetime of a rotating BH.

In most works on the subject, a simple power law parameterization of 
the disc emissivity $\epsilon(r) \propto r^{-\beta}$ is usually adopted. 
This is done also in the analysis of W01:
letting $\beta$ be a free fitting parameter, they find
$\beta \sim 4$, a value much larger than usually found
in Seyfert galaxies (Nandra et al., 1997). \\

In order to provide a physical picture of a so steep emissivity,
W01 invoke strong magnetic stresses acting in the innermost part
of the system, which dissipate a considerable amount of energy in the disc
at very small radii. If the magnetic field lines thread the BH horizon, 
this would imply magnetic extraction of the BH rotational energy --
the so called {\it Blandford-Znajek} effect (Blandford \& Znajek, 1977;
cited as BZ hereafter). However, the efficiency of the BZ effect 
has been questioned in recent years by e.g. Ghosh \& Abramowicz 
(1997) and Livio, Ogilvie \& Pringle (1999). 
These works argue that the electromagnetic output 
from the inner disc regions should in general dominate over 
that due to the BH. Thus the BH spin would probably be irrelevant 
to the expected electromagnetic power output from the system. 

Krolik (1999), Agol \& Krolik (2000) and 
Li (2000) proposed that MHD processes play a dominant role 
if magnetic field lines connect downfalling plasma near the
hole with more distant regions: high efficiency of energy 
extraction can be achieved in this way even if 
the magnetic field does not thread the horizon itself.
This magnetized accretion offers an alternative to the original BZ 
process and to its follow-up generalizations (e.g., Phinney 1983); 
however, the mechanism is violently non-stationary and such situations 
have not been quantitatively modelled yet (cf. Koide et al.,
2000, and Tomimatsu \& Takahashi, 2001, for the first attempts of such
modelling). \\

In this paper we show that the required steep emissivity law,
as well as the line EW and the amount of Compton reflection,
may be reproduced with a phenomenological model in which a X-ray 
illuminating source is located on the BH symmetry axis
({\it lamp-post} model: Martocchia \& Matt, 1996, Petrucci \& Henri, 
1997, Bao, Wiita \& Hadrava, 1998, Reynolds et al., 1999, 
Dabrowski \& Lasenby, 2001). This can be considered as a simplified
scheme, appropriate for various physical scenarios, including 
the mentioned MHD energy extraction. Indeed, Agol \& Krolik (2000) state
that magnetized accretion may also lead to enhanced coronal
activity immediately above the plunging region. ``If so,
this would provide a physical realization for models (...) which
call for a source of hard X-rays on the system axis a few
gravitational radii above the disc plane''.
Alternatively, shock waves in an aborted jet close to the BH axis 
have been proposed as a source of the central irradiation
by Henri \& Petrucci (1997). This model assumes that a point source 
of relativistic leptons (e$^+$,e$^-$) illuminates the accretion disk 
by Inverse Compton process; the resulting angular and spectral distribution 
of soft and hard radiation has been derived.

\section{A centrally illuminated disc in Kerr metric}

The primary source is schematically supposed to be pointlike and 
located at a height $h$ on the system symmetry axis. 
This allows to estimate the main effects resulting from various 
degrees of anisotropy of the illumination just by varying $h$.

If the primary source, assumed to be isotropic in its own reference
frame, is very close to the BH, a fraction of photons emitted towards
infinity are deflected by the BH gravitational field and
illuminate the accretion disc, at the same time
increasing the number of X-rays able to produce line emission
and reducing the primary radiation observed at infinity (Figure 
\ref{fig:solidan}; see Martocchia \& Matt, 1996, and Martocchia, 
2000, for details). For static BHs this effect is
counterbalanced by the loss of solid angle subtended by the matter
to the source when the latter is very close to the BH 
(because $r_{\rm{ms}}=6r_{\rm{g}}$). Martocchia \& Matt (1996) showed
that in the case of a spinning BH the increase in the line
intensity can be up to a few times the value for a static BH, and 
that the increase in the equivalent width can be even more dramatic. 
Due to the high (relativistic) orbital speed of the disc medium 
at very low radii, the impinging photons as seen in the matter's 
frame arrive with high incident angles, which further increases the 
local emissivity of fluorescence and reflection. The local emissivity 
of fluorescence is further increased by the blue-shift of these photons.
 
\bfg
\centering
\includegraphics[width=\columnwidth]{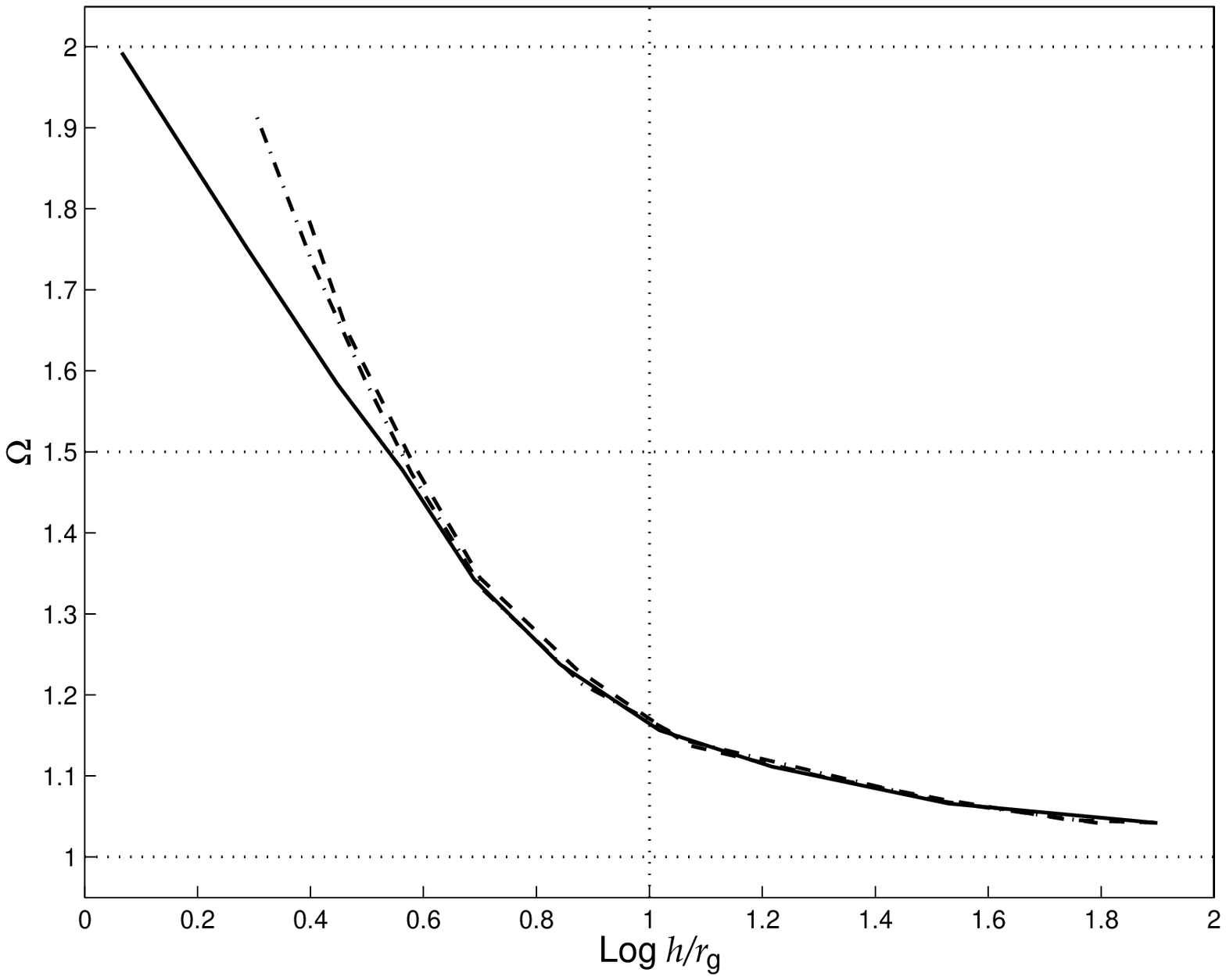}
\caption{$\Omega$ is
the ratio between the solid angle of the
radiation which reaches the equatorial plane
and the same solid angle as it would come out in a 
flat spacetime ($\Omega=\Omega_{\rm{disc}}/\Omega_{\rm{cl}}$), 
in the assumption that the disc extends 
up to $r_{\rm out}=1000r_{\rm{g}}$, and for
three values of the BH spin (solid line: $a/m=1.$; dot-dashed: $a/m=0.5$; 
dashed: static BH). 
The photon deflection towards the disc, i.e. the anisotropy
of the illuminating radiation field, strongly increases
with decreasing primary source height. }
\label{fig:solidan}
\efg 

Using the {\it lamp-post} framework, iron line profiles with the 
underlying Compton-reflected continuum have been presented in Martocchia, 
Karas \& Matt (2000). Monte Carlo simulations have been used to 
calculate the photon transfer within disc matter on the base of 
Compton scattering, self-consistently taking into account GR effects 
such as those on the illuminating photons' impinging angle. 

\subsection{Emissivity law}

In the {\it lamp-post} as well as in more realistic models,
the actual profile of the disc emissivity -- which depends on the 
geometry of the illuminating matter and is affected by many factors,
including general-relativistic (GR) effects -- is more complex than a
simple power-law. Appropriate emissivity laws $\epsilon(r)$ have been 
presented in Martocchia (2000) and Martocchia, Karas \& Matt (2000).

\bfg
\centering
\includegraphics[angle=0,width=\columnwidth]{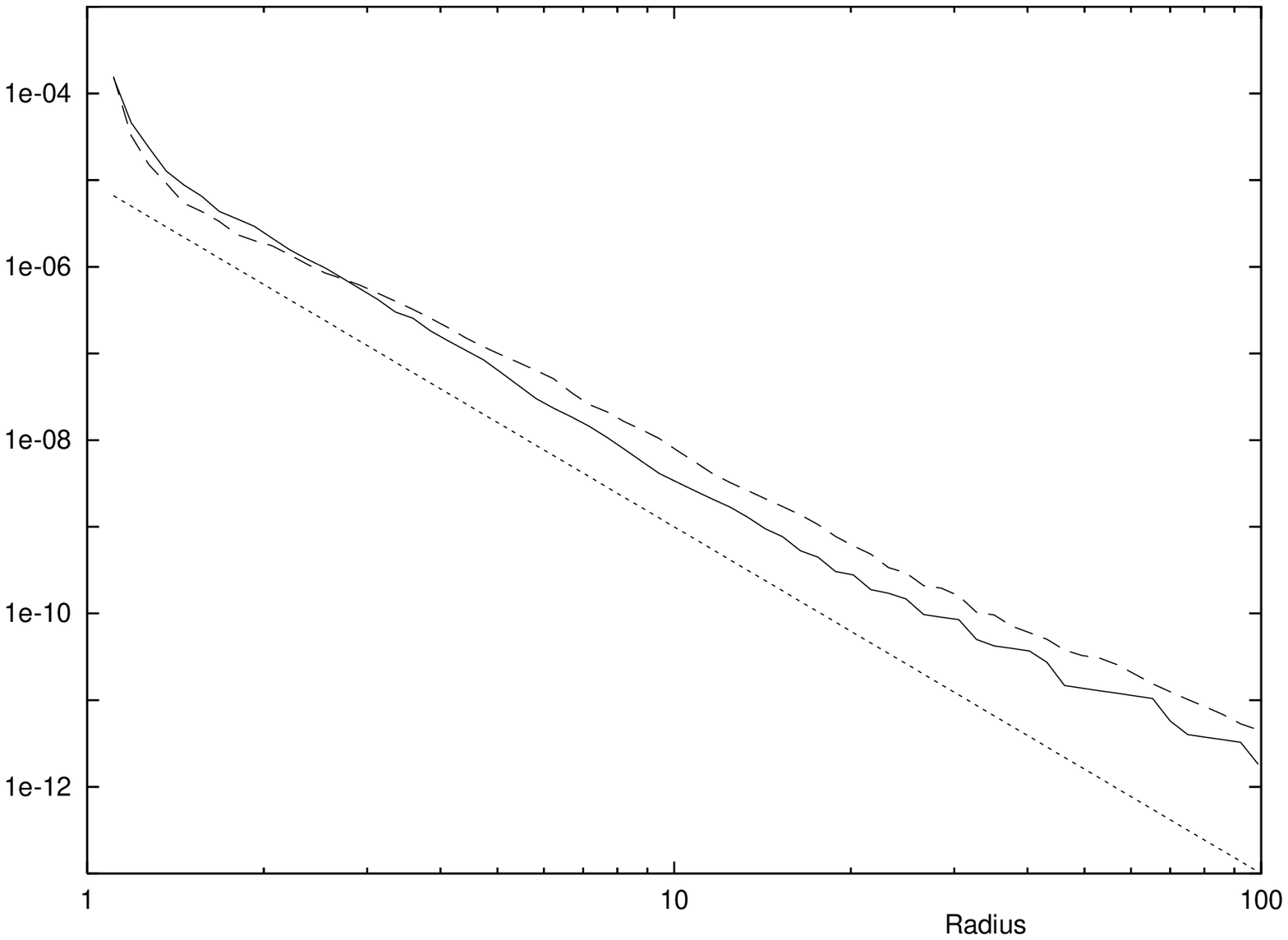}
\caption{Disc emissivity (in arbitrary units) vs. radius 
(in units of $r_{\rm{g}}$), for a source located at $h = 3r_{\rm{g}}$ 
(solid line) 
and $h = 4r_{\rm{g}}$ (dashed line) in the metric of a maximally 
spinning BH ($a\simeq0.9981m$). Lower, a straight 
dotted line corresponds to a power law with $\beta=-4$.
More plots, for different values of $h$, have been
presented in Martocchia, Karas \& Matt (2000).}
\label{fig:emiss}
\efg

In particular, $\epsilon(r)$
steepens when $h$ decreases, because of the enhanced anisotropy of 
the primary emission (see Figure~\ref{fig:emiss}). 
It can be approximated by functions of the form
$\epsilon(r) = A r^{-B} + C r^{-D}$.
The best-fit coefficients for this formula, obtained by least
square approximation, have been presented in
Martocchia, Karas \& Matt (2000).

\begin{figure}
\centering
\includegraphics[angle=-90,width=\columnwidth]{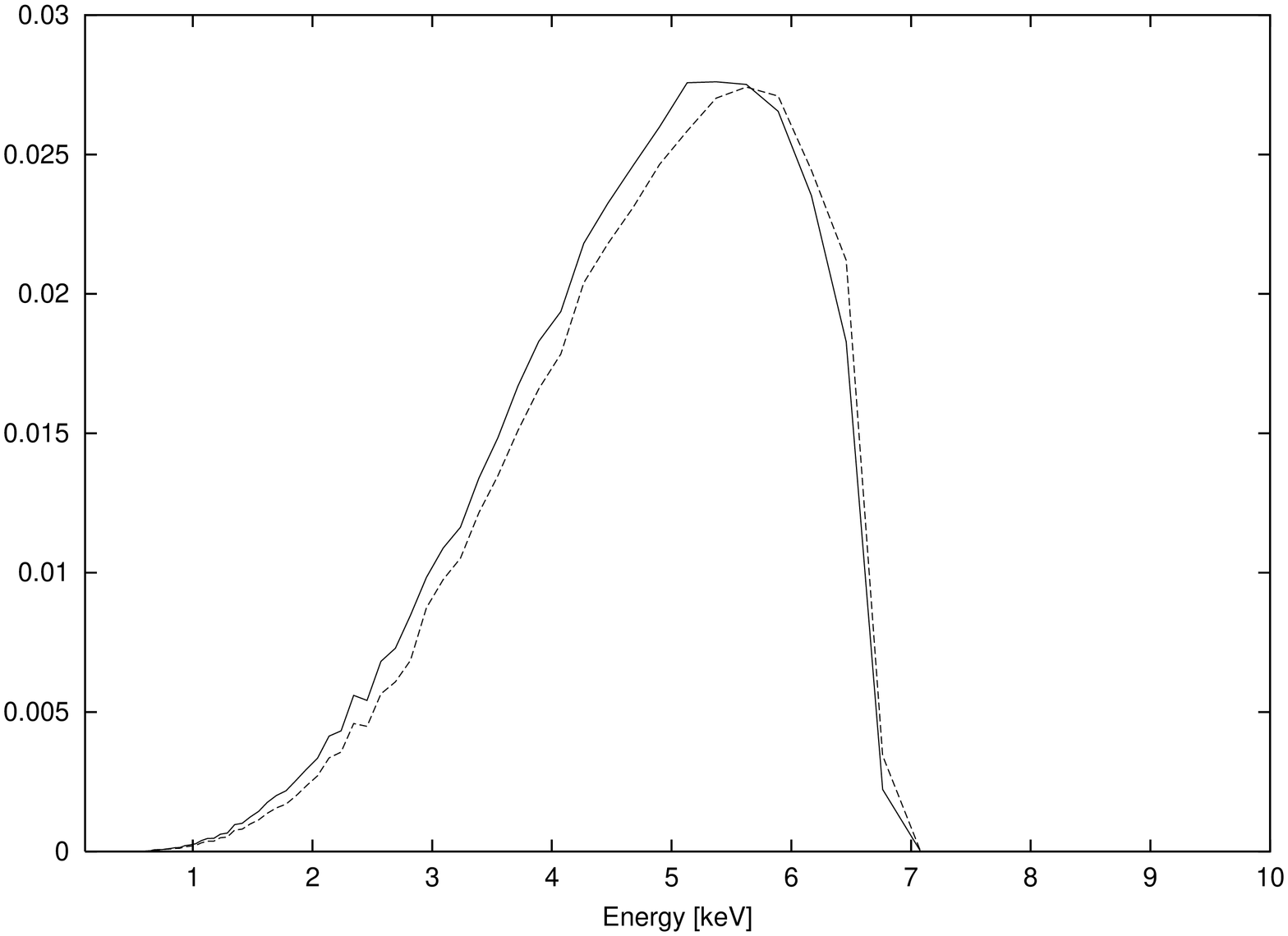}
\caption{Fe \ka\ profiles computed through the {\sc XSPEC} routine 
{\sc kerrspec} (Martocchia 2000), using respectively
a power-law emissivity with $\beta = 4$ (solid line)
and a lamp-post emissivity with $h=3r_{\rm{g}}$ (dashed line).
All other parameters have been fixed equal to the best-fit values 
reported by W01. Differences between the two models are clearly
too subtle to be detectable with {\it XMM}.}
\label{fig:models}
\end{figure}

With decreasing $h$, the effect of light bending is enhanced
(Martocchia \& Matt, 1996) and the fraction of (primary)
photons impinging onto the innermost regions of the disc increases.
It is easy to recognize 
(cf. Figure~\ref{fig:emiss}) that an emissivity law with $\beta \sim 4$,
i.e. the one derived by W01, may be produced in
our model with a small height of the primary source 
($h \sim 3\div4r_{\rm{g}}$).
For an emissivity law of this kind, the line profile 
comes out to be very broad and red-shifted (cf. Figure~\ref{fig:models};
see Martocchia, Karas \& Matt, 2000, for details). The differences
between this profile and that obtained by a powerlaw emissivity
with $\beta \sim 4$ are clearly too small to be measurable by {\it XMM}.

\subsection{Line EW and reflection continuum }

In the previous section we have shown that an emissivity law 
similar to the one derived by W01 in the {\it XMM-Newton} spectrum
of MCG--6-30-15 may be obtained in the framework of the
simple {\it lamp-post} model. Here we would like to stress that
the large observed EW and $R$ may be also self-consistently reproduced.

Martocchia \& Matt (1996) showed that the model predicts 
an anti-correlation between the intensity
of the reflected features (line and continuum) and the intensity 
of the primary flux, assuming that the latter is due to a change 
in the average height of the emitting matter.
Indeed, for a low primary source ($h \sim 3r_{\rm{g}}$) values of
EW$\sim 250$ eV or more may be easily obtained (e.g. Martocchia, Karas 
\& Matt, 2000.)
When allowing the source to be located off the axis of rotation, an 
even stronger enhancement can be obtained (Dabrowski \& Lasenby, 2001).

The {\it lamp-post} picture predicts that the 
intensity of the Compton-reflected continuum
correlates with the intensity of the fluorescent emission line.
The best-fit value for $R$ found by W01 is not well constrained,
but may be as high as $\sim7.6$.
In our model (see Figure~\ref{fig:solidan}),
a primary source located at $h \sim 3r_{\rm{g}}$ produces an anisotropy 
ratio $\sim1.6$ (thus $\Omega_{\rm disc} \sim 3.2\pi$), i.e. we find
$$R=\frac{\Omega_{\rm disc}}{4\pi-\Omega_{\rm disc}} \sim 4.$$

Let us finally determine the order of magnitude of the intrinsic 
luminosity of the point source in these assumptions. 
From the observed flux we can estimate the intrinsic (rest frame) 
value of the illumination in the 2--10 keV band by use of the 
anisotropy and gravitational redshift factors (the latter 
given by $g_{h=3m}\simeq0.63$, cp. Martocchia \& Matt 1996), 
and assuming an energy index $\alpha=1.3$,
$H_0\simeq 70$ km s$^{-1}$ Mpc$^{-1}$ and $z \simeq 0.008$, 
we get the following primary source luminosity:
$$L_{\rm{s}} \sim 4\pi D^2 F_{\rm{2-10~keV}} \times g_h^{1-\alpha} 
\times 1.6 \simeq 5.3 \times 10^{42} \rm{~erg~s}^{-1}.$$ \\

In conclusion, we have shown that a simple {\it lamp-post} model is able to
describe the results presented by W01 equally well and in a consistent way. 
It reproduces the observed emissivity and explains the large amount 
of line flux and reflection at the same time, 
providing that the primary X-ray source is 
located at $h \sim 3r_{\rm{g}}$.

\begin{acknowledgements}
AM and GM acknowledge financial support from 
MURST under grant {\sc cofin-00-02-36}, GM also from ASI,
VK from GACR 205/00/1685. We thank the referee, J. Wilms,
for his useful suggestions.
\end{acknowledgements}




\begin{thebibliography}{}
\bibitem[]{} Agol E. \& Krolik J.H., 2000, ApJ 528, 161
\bibitem[]{} Bao G., Wiita P. J. \& Hadrava P., 1998, ApJ 504, 58 
\bibitem[]{} Blandford R.D. \& Znajek R.L., 1977, MNRAS 179, 433 [BZ]
\bibitem[]{} Dabrowski Y. et al., 1997, MNRAS 288, L11 
\bibitem[]{} Dabrowski Y. \& Lasenby A.N., 2001, MNRAS 321, 605
\bibitem[]{} Fabian A.C., Iwasawa K., Reynolds C.S. \& Young A.J., 2000,
PASP 112, 1145
\bibitem[]{} Ghosh P. \& Abramowicz M.A., 1997, MNRAS 292, 887
\bibitem[]{} Guainazzi M. et al., 1999, \asa\ 341, L27
\bibitem[]{} Henri G. \& Petrucci P.O., 1997, \asa\ 326, 87
\bibitem[]{} Iwasawa K. et al., 1996, MNRAS 282, 1038 
\bibitem[]{} Koide S., Meier D.L., Shibata K., Kudoh T., 2000, \apj\ 536, 668
\bibitem[]{} Krolik J.H., 1999, \apj\ 515, L73
\bibitem[]{} Laor A., 1991, ApJ 376, 90 
\bibitem[]{} Li-Xin Li, 2000, \apj\ 540, L17
\bibitem[]{} Livio M., Ogilvie G.I. \& Pringle J.E., 1999, ApJ 512, 100
\bibitem[]{} Martocchia A., 2000, {\it ``X-ray Spectral Signatures of 
	Accreting Black Holes''}, PhD Thesis, SISSA-ISAS, Trieste
\bibitem[]{} Martocchia A., Karas V. \& Matt G., 2000, MNRAS 312, 817 
\bibitem[]{} Martocchia A. \& Matt G., 1996, MNRAS 282, L53 
\bibitem[]{} Nandra K. et al., 1997, ApJ 477, 602 
\bibitem[]{} Petrucci P.O. \& Henri G., 1997, \asa\ 326, 99 
\bibitem[]{} Phinney E.S., 1983, {\it ``A Theory of radio sources''}, 
	PhD thesis, Univ. Cambridge
\bibitem[]{} Reynolds C.S. \& Begelman M.C., 1997, ApJ 487, 109
\bibitem[]{} Reynolds C.S. et al., 1999, ApJ 514, 164 
	Comput. Phys. Commun. 88, 109
\bibitem[]{} Sako M. et al., 2001, \apj, submitted - {\sf astro-ph}/0112436
\bibitem[]{} Tanaka Y. et al., 1995, Nat. 375, 659 
\bibitem[]{} Thorne K.S., 1974, \apj\ 191, 507 
\bibitem[]{} Tomimatsu A. \& Takahashi M., 2001, \apj\ 552, 710
\bibitem[]{} Wilms J., Reynolds C.S., Begelman M.C., Reeves J., Molendi S.,
Staubert R., Kendziorra E., 2001, MNRAS, 328, L27 [W01]
\end{thebibliography}
\end{document}